\documentclass[aps,pra,reprint,amsmath,amssymb,superscriptaddress]{revtex4-1}
\usepackage{graphicx}
\usepackage{mathrsfs}
\renewcommand{\phi}{\varphi}
\usepackage{xfrac}
\usepackage[colorlinks]{hyperref}
\usepackage{mathrsfs}
\usepackage{times}
\usepackage[utf8]{inputenc}
\mathchardef\Re="023C
\mathchardef\Im="023D
\usepackage{subcaption}
\usepackage{slashed}

\usepackage{color}

\makeindex
\begin{document}

\title{Rayleigh Edge Waves in Two-Dimensional Crystals with Lorentz Forces \newline - from Skyrmion Crystals to Gyroscopic Media}

\author{Claudio Benzoni}
\thanks{C.B. and B.J. contributed equally to this work.}
\affiliation{Physik-Department, Technische Universität München, 85748 Garching, Germany}
\affiliation{Munich Center for Quantum Science and Technology (MCQST), 80799 München, Germany}
\author{Bhilahari Jeevanesan}
\thanks{C.B. and B.J. contributed equally to this work.}
\affiliation{Physik-Department, Technische Universität München, 85748 Garching, Germany}
\affiliation{Munich Center for Quantum Science and Technology (MCQST), 80799 München, Germany}
\author{Sergej Moroz}
\affiliation{Physik-Department, Technische Universität München, 85748 Garching, Germany}
\affiliation{Munich Center for Quantum Science and Technology (MCQST), 80799 München, Germany}
\begin{abstract}

We investigate, within the framework of linear elasticity theory, edge Rayleigh waves of a two-dimensional
elastic solid with broken time-reversal and parity symmetries due to a Berry term. As our prime example, we study the elastic edge wave traveling along the boundary of a two-dimensional skyrmion lattice hosted inside a thin-film chiral magnet.
We find that the direction of propagation of the Rayleigh modes is determined not only by the chirality of the thin-film, but also by the Poisson ratio of the crystal. We discover three qualitatively different regions distinguished by the chirality of the low-frequency edge waves, and study their properties. To illustrate the Rayleigh edge waves in real time, we have carried out finite-difference simulations of the model.  Apart from skyrmion crystals, our results are also applicable to edge waves of gyroelastic media and screened Wigner crystals in magnetic fields.
Our work opens a pathway towards controlled manipulation of elastic signals along boundaries of crystals with broken time-reversal symmetry.
\end{abstract}

\maketitle

\section{Introduction}
Recent years have seen a new surge of excitement around chiral surface waves in hydrodynamics \cite{Delplace1075, PhysRevX.7.031039, 2018abanov, 2019wiegmann, PhysRevLett.122.128001, 2019abanov, PhysRevResearch.2.013147, graf2020topology}. 
The role of bulk topology for the existence and robustness of such waves has been vigorously investigated \cite{Delplace1075, PhysRevX.7.031039, PhysRevLett.122.128001, PhysRevResearch.2.013147, graf2020topology}. Chiral surface modes have been also recently used as a tool to measure the bulk Hall viscosity of an active fluid \cite{soni2019odd}. 

In this work, we study surface waves in two-dimensional crystals which break parity $P$ (spatial reflection) and time-reversal $T$ symmetry, but preserve the combined $PT$ symmetry.  In the quantum realm, well-known examples of such systems are two-dimensional thin-film chiral magnets, which host lattices formed out of skyrmion defects \cite{muhlbauer2009skyrmion, yu2010real, yu2011near, han2017skyrmions}, Wigner crystals in a magnetic field \cite{Fukuyama} and Abrikosov vortex lattices in superconductors and rotating superfluids \cite{Sonin2016}. In the last few years such crystals were also designed in gyroscopic metamaterials \cite{Nash14495}, \cite{phononic_gyroscopes} and mass-spring networks subject to Coriolis forces \cite{designed_mechanical}.

The investigation of waves that propagate along the free surface of an elastic solid, and whose disturbance remains confined to the vicinity of the boundary is an old topic that goes back to the remarkable classic paper by Rayleigh \cite{Rayleigh1885}, where an approximate numerical solution for the dispersion relation of such waves was obtained. Within linear elasticity theory, this excitation - known today as the Rayleigh wave - is non-dispersive and has speed lower than the bulk transverse and longitudinal sounds \cite{landauElasticity, thorne}.  As already anticipated by Rayleigh, these surface waves play a crucial role in  seismology \cite{stein2009introduction}.

\begin{figure*}[t]
\includegraphics[width=2\columnwidth]{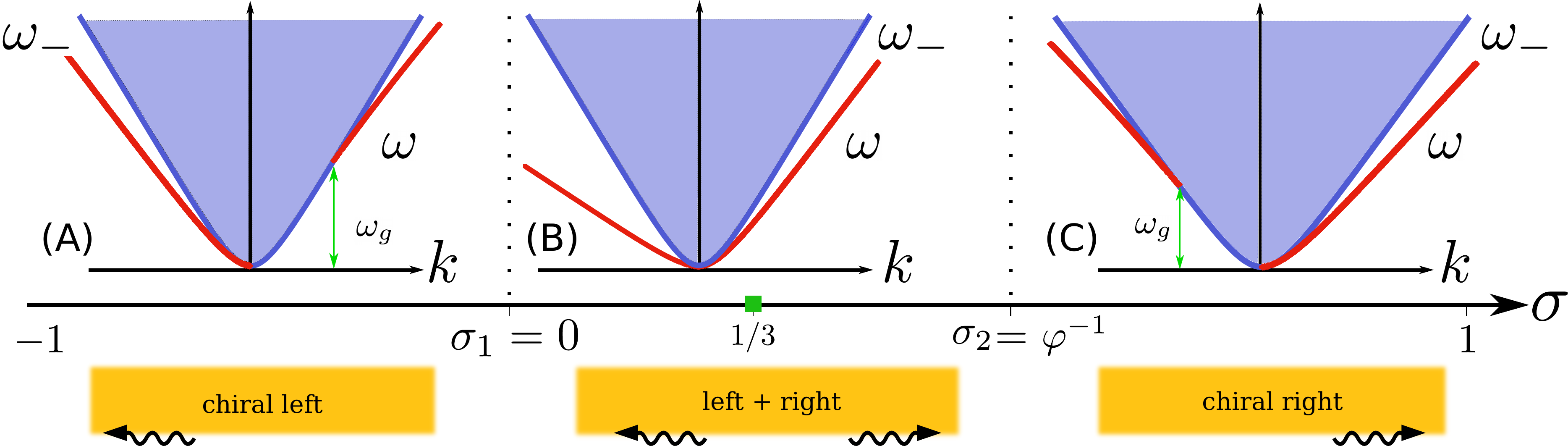} 
\caption{Sketched 	dispersion relations of Rayleigh surface excitations $\omega(k)$ (red) as a function of the Poisson ratio $\sigma$. In cases (A) and (C) the low-frequency spectrum is chiral, while in the intermediate regime $0 \leq \sigma \leq \varphi^{-1}$ edge waves of both chiralities are present.
Here $\varphi \equiv (1+\sqrt{5})/2$ is the golden ratio. 
The green square denotes the point $\sigma = 1/3$ where the edge wave spectrum is symmetric. Plotted in blue are cross sections of the gapless bulk excitation.}
\label{Fig:regimes}
\end{figure*}

The general focus of this paper is the investigation of a long-wavelength effective field theory of a two-dimensional skyrmion lattice, where the Cartesian components of the displacement from equilibrium positions $u^x$ and $u^y$  are coupled by a Berry term \cite{petrova2011spin, PhysRevLett.107.136804}. The displacements are assumed to be small, which allows the framework of linear elasticity to be employed. 
We find that the behavior of the edge-waves can be tuned by changing the Poisson ratio $\sigma$ \cite{chaikin1995principles}. In fact, we show that there exist three qualitatively distinct phases, captured by the diagram in Fig.  \ref{Fig:regimes}. The phases are distinguished by the propagation direction of their low-frequency surface waves. In the long-wavelength and low-frequency limit we develop an analytic treatment of these edge waves.

\section{Skyrmion crystal elasticity in thin-film chiral magnets} 
It is well-known that an elementary skyrmion defect in a ferromagnet experiences an effective magnetic field $\mathcal{B}$ and the associated Lorentz force because it picks up the Berry phase of $2\pi$ whenever encircling a spin $1/2$  \cite{PhysRevB.53.16573}. Moreover, a skyrmion can be characterized by a finite inertial mass $m$, which was derived in \cite{makhfudz2012inertia} by integrating out fluctuations of its spatial profile.
In this paper we study the surface waves of two-dimensional skyrmion lattices present in thin-film chiral magnets such as $\text{Fe}_{0.5}\text{Co}_{0.5}\text{Si}$ \cite{yu2010real} and $\text{FeGe}$ \cite{yu2011near}. Starting from the continuum theory of \cite{petrova2011spin, PhysRevLett.107.136804}, the skyrmion dynamics is described by a field theory of coarse-grained elastic variables $u^i \left(\mathbf{x}\right)$  with $i= x,y$, denoting the displacements of skyrmions from their equilibrium positions, see \cite{han2017skyrmions} for a pedagogical exposition. 
The action of the skyrmion displacement field is given by
\begin{equation}\label{Lagrangian}
S\left[u^i \right]=  \int dt~ d^2 x \left[   \frac{\rho}{2} \dot{u}^2 - \frac{\rho \Omega}{2} \epsilon_{ij}u^i \dot{u}^j   - \mathcal{E}_{{\rm el}}\left( u_{ij} \right) \right],
\end{equation}
where the overdot denotes the time derivative, $\rho$ is the mass density of the skyrmions,  $\Omega=\mathcal{B}/m$ is the cyclotron frequency associated with effective magnetic field $\mathcal{B}$, $u_{ij}\equiv \partial_{\left( i\right.} u_{\left. j \right)}$ is the symmetric linearized strain tensor, and $\mathcal{E}_{{\rm el}}\left( u_{ij} \right)$ the elastic energy density, dictated by the geometry of the crystal. The convention for the completely antisymmetric Levi-Civita symbol is $\epsilon_{xy}=-\epsilon_{yx}=1 $ and summation over repeated indices is understood. At low frequencies, the Berry term in Eq. \eqref{Lagrangian}, which gives rise to an effective Lorentz force, dominates the first term that encodes Newtonian dynamics. As a result, $u^x$ and $u^y$ form a canonically conjugate pair of variables in the limit $m\to 0$. The Berry term breaks the time-reversal $T$  ($t\to-t$) and parity $P$ ($x\to -x$, $u^x\to - u^x$) symmetries, but preserves their combination.\newline
We assume that skyrmions form a triangular lattice, whose symmetry class, $C_6$, limits $\mathcal{E}_{{\rm el}}\left( u_{ij} \right)$ in two dimensions to the isotropic form \cite{landauElasticity, chaikin1995principles} 
\begin{equation}\label{elasticenergy}
\mathcal{E}_{{\rm el }}\left(u_{ij}\right) =  2C_1 u_{kk}^2+ 2 C_2  \tilde{u}_{ij}^2,
\end{equation}
where $\tilde{u}_{ij} \equiv u_{ij}-\left( u_{kk}\delta_{ij}\right)/2$ is the traceless symmetric part of the strain tensor. The \textit{compressional} elastic modulus $ 2 C_1$ quantifies the change of energy due to deformations that preserve the shape of the system but change its volume, while the \textit{shear} modulus $C_2$ fixes the energy cost of volume-preserving deformations. In the context of effective field theories, $C_1$ and $C_2$ are just parameters of the derivative expansion and can take arbitrary non-negative values \footnote{For more complicated systems in which the elastic displacements $u^i$ are coupled to other degrees of freedom, as for example in vortex crystals \cite{baym2003tkachenko}, the compressional modulus $C_1$ can be negative.}. 

While the action in \eqref{Lagrangian} is the effective theory governing skyrmion lattice dynamics, at long wavelengths it also describes elastic gyroscopic systems in the limit of small nutation angle \cite{garau2019transient}. Moreover, it can also be used to describe certain screened Wigner crystals in magnetic fields \cite{edge_vignale}. 

The action \eqref{Lagrangian} is quadratic and  the resulting equations of motion are linear
\begin{equation}\label{EOM}
\ddot{u}^i+\Omega \epsilon_{ij}\dot{u}^j=2 v_1\partial_i \left(\partial_p u^p\right)+v_2\partial^2 u^i,
\end{equation}
where we introduced $v_1 \equiv 2C_1 / \rho$, $v_2 \equiv 2C_2/\rho$. 
Assuming an infinite system that respects magnetic translational invariance in both directions, the equations of motion (\ref{EOM}) are algebraic in frequency/wavevector space.  Contrary to the situation where the Berry term is absent, the modes do not decouple into the longitudinal and transverse components. 
The set of equations \eqref{EOM} can be solved with elementary methods, and the two solutions for the dispersion relations are given by 
\begin{equation}\label{general_dispersion}
\omega_{\mp}^2  = \frac{\Omega^2}{2}+ (v_1+v_2)k^2 \mp k^2 \sqrt{v_1^2+\left(v_1+v_2\right)\frac{\Omega^2}{k^2} +\frac{\Omega^4}{4k^4}}.
\end{equation}
Due to their cumbersome form, the polarizations $\epsilon_{\pm}\equiv u^x_{\pm}/u^y_{\pm}$ are not presented in the general case here.
In this problem we can identify two distinct physical regimes:
(i) for small wavevectors/large magnetic fields the negative branch of Eq. \eqref{general_dispersion} gives rise to 
a gapless 
\emph{magnetophonon} mode with the quadratic dispersion
\begin{equation}\label{tk}
\omega_- \left(k\right) = \frac{\sqrt{v_2 \left(2v_1+v_2\right)}}{ \Omega }k^2 \left[ 1+O \left(\frac{k^2}{\Omega^2}\right) \right],
\end{equation}
 while the positive branch represents a gapped \emph{magnetoplasmon}  mode, dispersing as
 \begin{equation}\label{kohn}
\omega_+ \left(k\right) = \Omega \left[ 1+ \left(\frac{v_1+v_2}{\Omega^2}\right)k^2 + O \left(\frac{k^4}{\Omega^4}\right)\right].
\end{equation}
The latter mode is guaranteed to have the gap $\omega=\Omega$ at $k=0$ by the Kohn theorem \cite{Kohn}; the  system is analogous to a collection of single-species charged particles in a uniform magnetic field that interact through a potential which depends only on their relative distances.
In the zero wavevector limit, the polarization of the Kohn mode  is circular and its chirality is fixed by the sign of the effective magnetic field $\mathcal{B}$.
(ii) In the limit of large wavevectors, the Newtonian term dominates over the Berry term in Eq. \eqref{EOM} and  we asymptotically recover two linearly dispersing sound modes of a time-reversal invariant two-dimensional solid \footnote{In the absence of the Berry term the system supports a transverse and a longitudinal sound modes with respective group velocities $c_t^2 \equiv 2 C_2/\rho$  and $c_l^2 \equiv \left(4C_1+2C_2\right)/\rho$.}. In particular, at large momenta, the magnetophonon (\ref{tk}) merges into the transverse ($k_i u_-^i =0$) mode dispersing as $\omega_- = c_t k \left[  1+ O\left(\Omega^2/k^2\right) \right]$,  while the magnetoplasmon (\ref{kohn}) becomes the longitudinally polarized ($\epsilon_{ij}k_i u_+^j=0$) mode with $\omega_+ = c_l k\left[ 1+ O \left(\Omega^2/k^2\right) \right]$.

The Lagrangian naturally fits into a derivative expansion within the power-counting scheme $u^i \sim  O \left(1\right)$, $\partial_i \sim O \left( \varepsilon \right)$ , $\partial_t \sim O \left(\varepsilon^2\right)$
where we introduced a small parameter $\varepsilon \ll 1$. The difference in the power-counting of temporal and spatial derivatives originates from the soft quadratic dispersion of the magnetophonon.
All terms in the Lagrangian \eqref{Lagrangian}, except for the Newtonian term $\rho \dot{u}^2/2$, are of order $O(\varepsilon^2)$, defining the leading-order (LO) Lagrangian. On the other hand, the Newtonian term  scales as $\varepsilon^4$ and is less relevant at low frequencies, and thus is of the next-to-leading order (NLO). The inclusion of this term allows us to establish the crossover of edge waves that exist in the chiral system to the ordinary Rayleigh waves in the absence of a Berry term. We notice here that other NLO terms such as second-order elasticity $\tilde \lambda_{ijklmn}\partial_i \partial_j u_k \partial_k \partial_m u_n$ or the dissipationless phonon Hall viscosity  $\eta_{ijkl}\partial_i u_j \partial_k \dot{u}^l$ \cite{barkeshli2012dissipationless} are not included in the present work. 

In elastic media, internal stresses and forces are encoded in the stress tensor $T_{ij}= \delta \mathcal{E_{{\rm el }}} / \delta u_{ij}$ \cite{landauElasticity, thorne}. For a two-dimensional triangular crystal with the elastic energy density \eqref{elasticenergy} the stress tensor is
\begin{equation}\label{Tij}
T_{ij} = 4 C_1 u_{kk} \delta_{ij}+ 4 C_2 \tilde{u}_{ij}.
\end{equation}

\section{Rayleigh edge modes} We next turn to the study of exponentially localized Rayleigh waves that propagate on the edge of the skyrmion lattice. The breaking of time-reversal and parity symmetries in (\ref{Lagrangian}) due to the Berry term suggests that such modes might be chiral, i.e., propagating only in one direction, see e.g. \cite{1997monarkha} and \cite{edge_vignale}.
For the sake of simplicity, we consider the skyrmion crystal to fill the lower half-space with $y <0$. Without loss of generality we also choose $\Omega> 0$ throughout the rest of the paper. 

The translational invariance in time and along the horizontal direction motivates the ansatz $\textbf{u} \left(x,y, t\right)=   \textbf{u}~  e^{i \left(kx-\omega t\right)}e^{\kappa y}$ for a solution of (\ref{EOM}).
The wavevector along the boundary $k$ and the frequency $\omega$ are assumed to be real; confinement near the edge of the system requires the real part of $\kappa$  to be positive.

First, in order to make the following calculation more transparent, we shall focus on the low-frequency limit and drop the NLO Newtonian term $\sim \dot u^2$ in the model \eqref{Lagrangian}. The edge ansatz inserted into Eq. (\ref{EOM}) results in a characteristic equation for $\kappa$ with two solutions 
\begin{equation}\label{kappas}
\kappa_{1,2}(k,\omega) = \sqrt{k^2 \pm \frac{\Omega }{\sqrt{v_2(2 v_1  +v_2)}} \omega}+\mathcal{O}(k^2)
\end{equation}
 The corresponding eigenvectors $\textbf{u}_{1,2}\left(k,\omega \right)$ are functions of the wavevector $k$ and frequency $\omega$. Interestingly, here, in contrast to the ordinary Rayleigh construction, both solutions $\kappa_{1,2}$ originate from the single magnetophonon branch.

The general solution with given $k$ and $\omega$ is obtained by forming a linear superposition of $\textbf{u}_{1,2}\left(k,\omega \right)$ with two complex constants $a$, $b$
\begin{equation}\label{superposition}
\textbf{u}\left(x,y,t\right) = e^{i \left(kx-\omega t\right)} (a~\textbf{u}_{1} e^{\kappa_{1} y}+b~\textbf{u}_{2} e^{\kappa_{2} y}).
\end{equation}
Due to the $PT$ symmetry of the model, the dispersion satisfies $\omega(k)=- \omega(-k)$, hence it is sufficient to study only the interval $\omega\ge 0$. 

Here we will assume that the crystal is free at the boundary $y=0$. In this case there are no macroscopic forces acting on it from the outside. Thus there is no flux of linear momentum across the boundary surface at $y=0$, resulting in the so-called stress-free boundary conditions \footnote{Investigations of more general boundary conditions originating, for example, from the helical texture near the edge are deferred to a future work.}
\begin{equation}\label{BC}
T_{xy}(x,y=0) = T_{yy}(x,y=0)\stackrel{!}{=}0.
\end{equation} 
Substituting the ansatz \eqref{superposition} into the boundary conditions (\ref{BC}) results in the linear system of equations for $a$ and $b$
\begin{equation}\label{BCS}
\left(
\begin{array}{c c}
i k \sigma \epsilon_{1}+ \kappa_{1}  & ik\sigma \epsilon_{2} + \kappa_{2}\\
ik +\kappa_{1} \epsilon_{1}& ik+\kappa_{2}\epsilon_{2}\\
\end{array}
\right)\left(
\begin{array}{c}
a\\
b\\
\end{array}
\right)=0,
\end{equation}
where we have introduced the two-dimensional Poisson ratio $\sigma \equiv (2C_1 - C_2)/(2C_1 + C_2)$ and the shorthand $\epsilon_{1,2}\equiv u^x_{1,2}/u^y_{1,2}$. The dispersion relation $\omega(k)$ for the edge waves is obtained from the characteristic equation for the matrix in Eq. (\ref{BCS}). 

In traditional treatments of elasticity theory, the Poisson ratio was usually assumed to be a positive value \cite{landauElasticity}; however, in recent years, it was found that elastic systems can be engineered to have a negative Poisson ratio \cite{auxetic1, LAKES1038} and even more remarkably that such materials actually occur in nature \cite{yeganeh1992elasticity}. By now an explosion of research into exotic metamaterials has taken place (for an overview see \cite{auxetic3}), which go by the name {\it auxetic materials}. These systems have the counter-intuitive property that under uniaxial compression, they contract in the orthogonal direction. In the following we investigate the interval $-1 \leq \sigma \le 1$, where the elastic system is stable.

Substitution of the two edge modes into (\ref{BCS}) yields a dispersion relation $\omega(k)$ of the form
\begin{equation}\label{Disp}
\omega(k)=\alpha \frac{\sqrt{v_2(2 v_1 + v_2)}}{\Omega}  k^2
\end{equation}
with $\alpha$ being a non-negative and real solution of an unwieldy algebraic equation, which we investigate in detail in Appendix \ref{AppA}. This equation does not depend on the magnitude of $k$, but only on its sign. As a consequence, the equations for positive and negative $k$ are in general different, resulting in different solutions $\alpha(\text{sign}(k), \sigma)$, which we will denote by $\alpha_\pm(\sigma)$.  

\begin{figure}
\centering
\includegraphics[width=0.85\columnwidth]{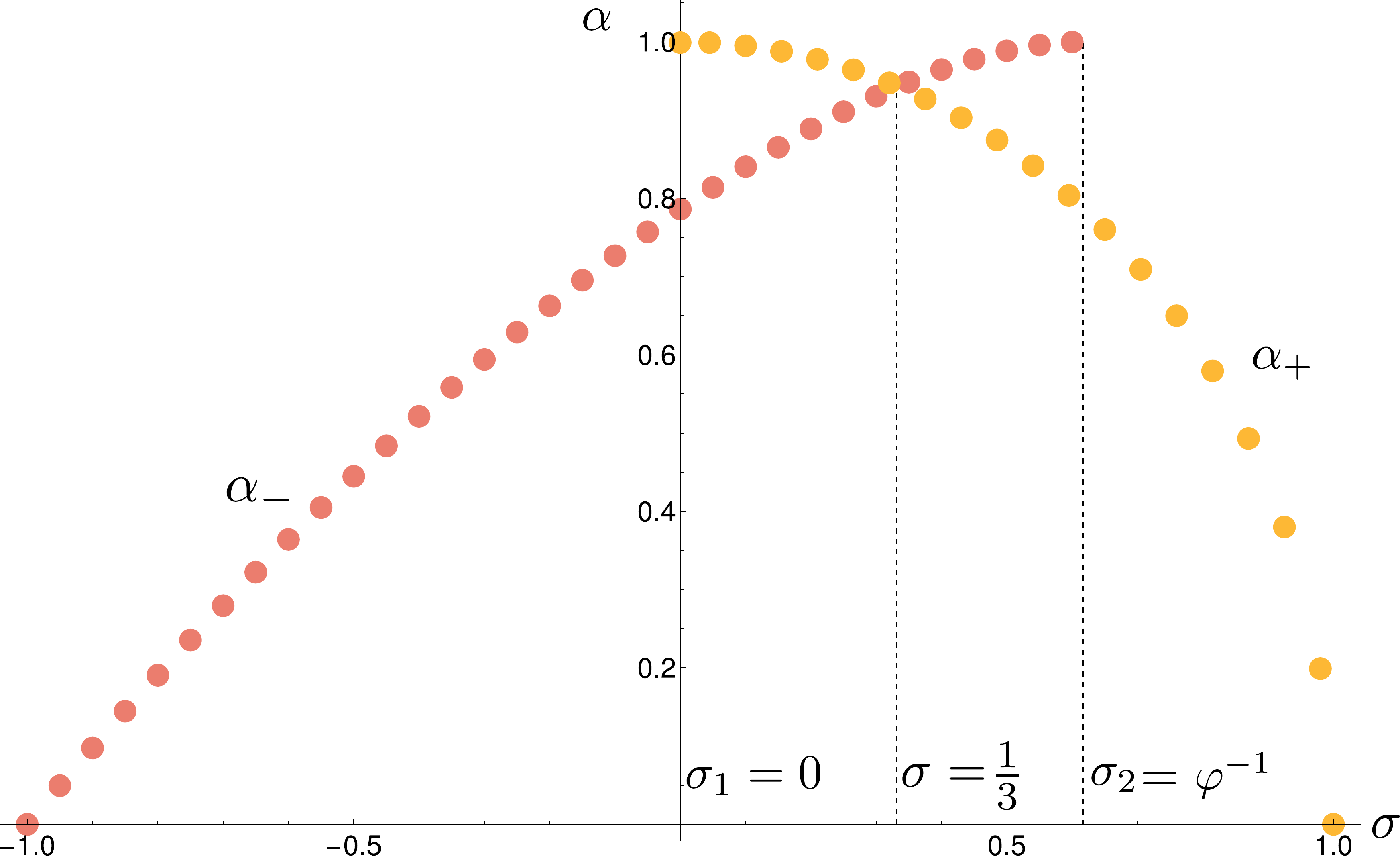} 
\caption{For small momenta the dispersion relation has the quadratic form $\omega =\alpha {\sqrt{v_2(2 v_1 + v_2)}}/{\Omega} \ k^2$. When only one of the branches $\alpha_\pm$ exists, the edge wave propagates unidirectionally. This happens for $\sigma < 0$ and $\sigma > \varphi^{-1}$, see Appendix \ref{AppA}.}
\label{Fig:alpha}
\end{figure}

We show the numerical solution of $\alpha_\pm(\sigma)$ in Fig. \ref{Fig:alpha}. 
As the value of $\sigma$ is varied, one finds three qualitatively different regimes. For $\sigma<0$ only the $\alpha_-$ branch exists: edge waves  can only propagate towards left, while propagation to the right is forbidden. We find analytically in Appendix \ref{AppA} that for $\sigma > \varphi^{-1}= (\sqrt{5}-1)/2$, i.e. the inverse golden ratio, the edge waves are once again chiral, but with propagation in the opposite direction. In the interval $0 \leq \sigma \leq \varphi^{-1}$ both branches of $\alpha_\pm$ exist and consequently edge waves can propagate in both directions.
The dispersion of the surface waves is generically asymmetric, since in general $\alpha_+ \neq \alpha_-$. However, it is clear from Fig.  \ref{Fig:alpha}, that at the point $\sigma= 1/3$ \footnote{A microscopic model that realizes the value $\sigma = 1/3$ is the triangular lattice of equal masses connected through nearest neighbours identical springs.} the spectrum is symmetric, see Appendix \ref{AppA} for the analytical justification, where we also determine the value $\alpha_\pm = {2 \sqrt{2}}/{3}$.

In the presence of the sub-leading Newtonian term we solved the edge problem numerically. The resulting spectrum is sketched in Fig. \ref{Fig:regimes}.  The inclusion of the Newtonian term results in propagation in a forbidden direction for momentum and frequency larger than a critical value $k_\text{crit}$ and $\omega_g \equiv \omega(k_\text{crit})$. We checked that ordinary Rayleigh waves are recovered for $\omega \gg \omega_g$.

In order to illustrate our findings, we have carried out finite-difference simulations of dynamics encoded by \eqref{EOM} subject to the boundary conditions (\ref{BC}) on a $500\times 500$ spatial grid; for details see Appendix \ref{AppB}. Fig. \ref{Fig:simulations} shows simulation snapshots for different values of the Poisson ratio. The initial condition for the displacement field is identical in all three simulations: the elastic medium has zero displacement everywhere, except for a small central region near the lower horizontal boundary, where it is deformed. Starting with this condition, we let the system evolve over time (see \cite{supmat} for a simulation video). We observe that while the excitation decays partially into the bulk of the medium, some part remains localized near the edge and travels along the boundary. For $\sigma = -0.8$ and $\sigma = +0.8$ one sees clearly how the edge excitations travel unidirectionally and in opposite directions for the two Poisson ratios. For $\sigma = 1/3$ we observe two edge excitations that travel symmetrically in both directions. 
\begin{figure}
\includegraphics[width=\columnwidth]{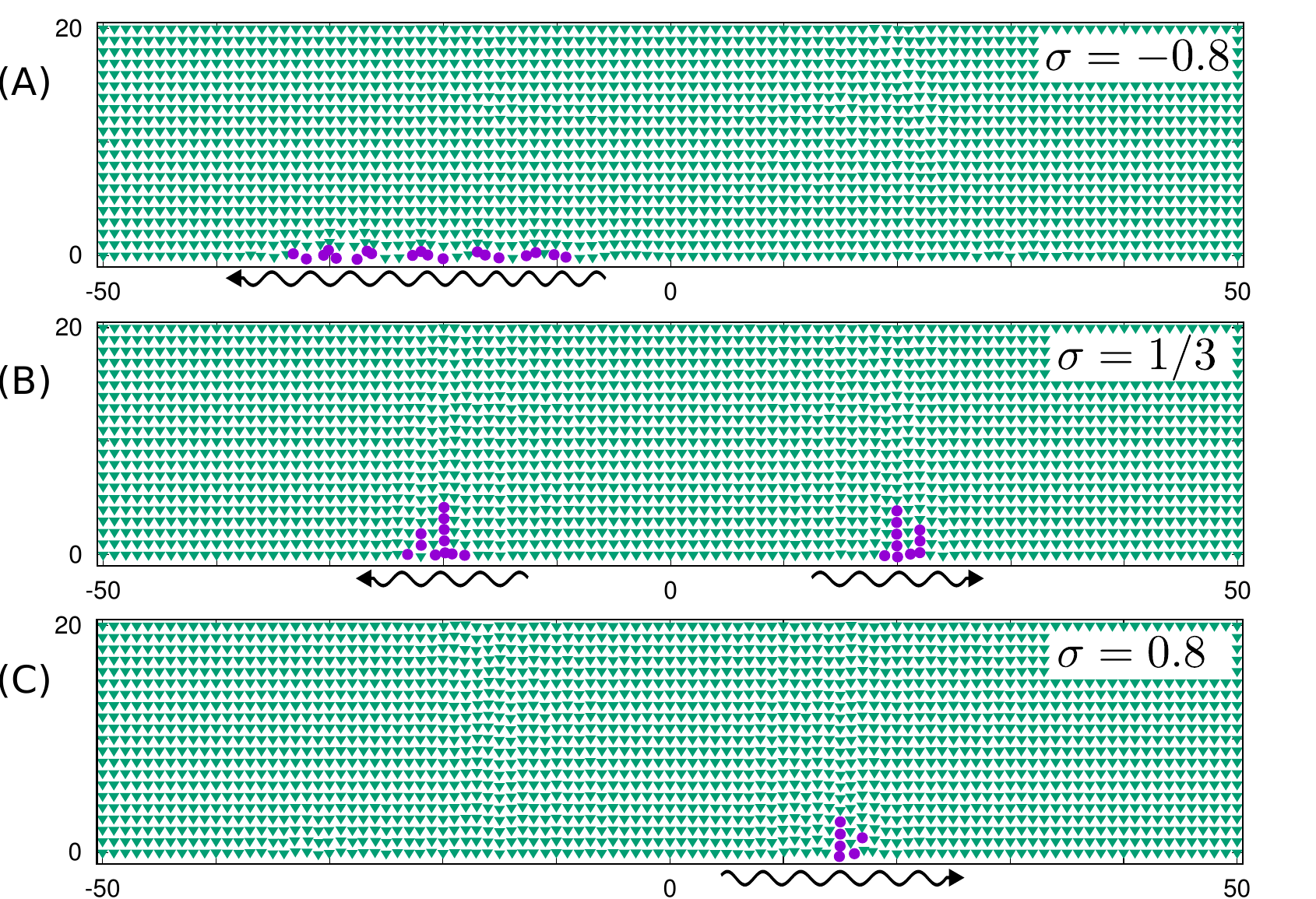} 
\caption{Edge excitations as seen in finite-difference simulations \cite{supmat}. In all three plots, the system was displaced in a small region near the boundary at $x = 0$ and evolved over time. The three values of $\sigma$ are representative of the three regimes shown in Fig. \ref{Fig:regimes}. To guide the eye, we colored in magenta the grid points that have large amplitude defined by a threshold value.}
\label{Fig:simulations}
\end{figure}

\begin{figure}
\includegraphics[width=\columnwidth]{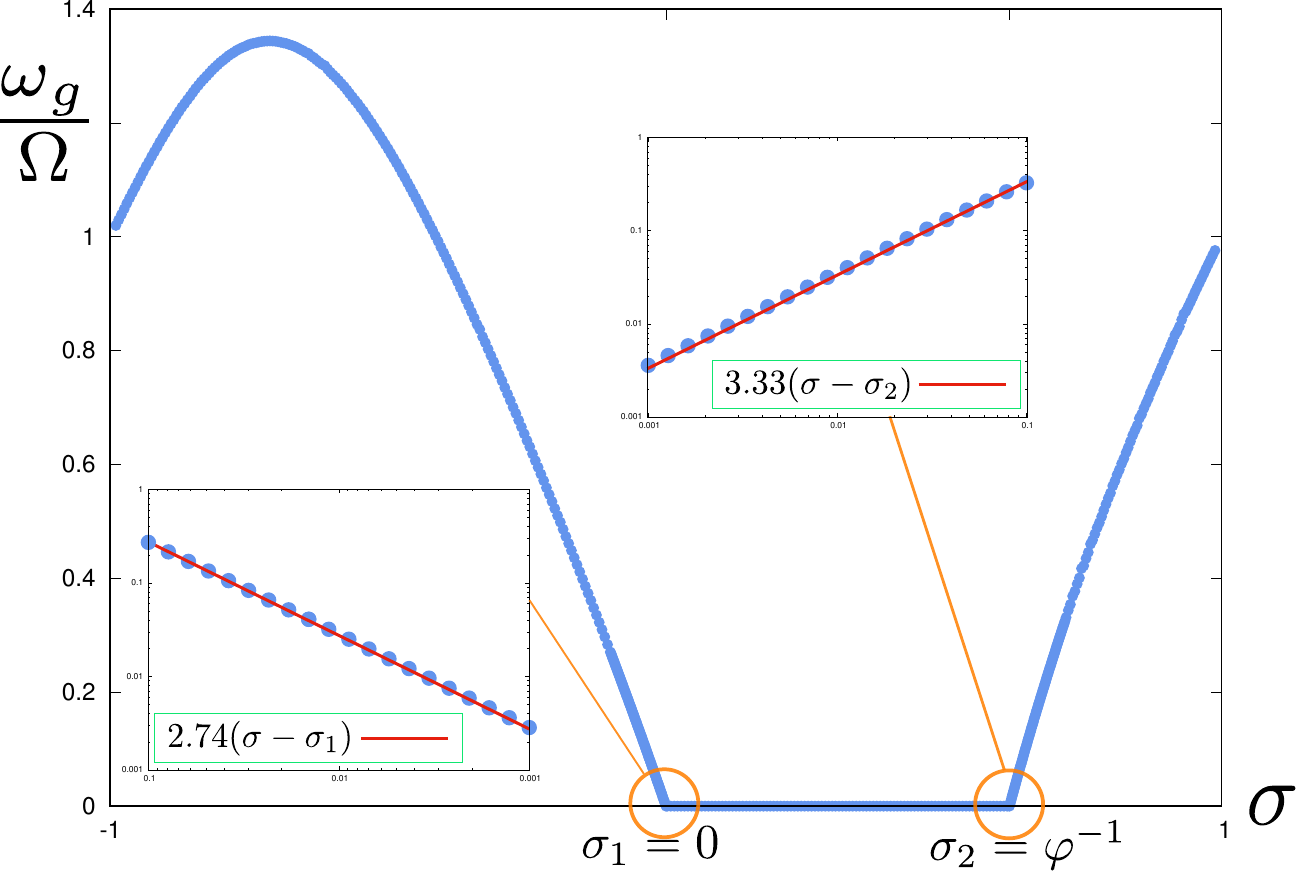}
\caption{Frequency gap $\omega_g$ of the edge waves as a function of the Poisson ratio $\sigma$. The gap is zero in the interval $[\sigma_1, \sigma_2]$. The insets show that the gap vanishes linearly near the critical points $\sigma_1$ and $\sigma_2$.}
\label{Fig:height}
\end{figure}

In order to investigate the transition between the  three regimes, we studied the magnitude of the frequency gap $\omega_g$ as a function of the Poisson ratio $\sigma$. The result is displayed in Fig. \ref{Fig:height}. 
The figure demonstrates that the non-chiral regime ($\omega_g=0$) exists inside a finite interval $\sigma_1<\sigma<\sigma_2$. This implies that the gap vanishes in a non-analytic way, reminiscent of the behavior of an order parameter near a continuous phase-transition. Indeed, we find that the gap $\omega_g$ vanishes linearly near the critical ratios $\sigma_1=0$ and $\sigma_2=\varphi^{-1}$, see the insets of Fig. \ref{Fig:height}.


A particularly simple case of surface modes is found in the limit where the compressional modulus $C_1$ vanishes, i.e. for $\sigma=-1$. In the time-reversal invariant setting, this {\it maximally auxetic} problem emerges in the twisted  Kagome lattice \citep{sun2012surface}. 
We find for our system that at $\sigma = -1$ edge modes exist and the frequency spectrum is a flat band. This implies that once a deformation is introduced at the edge of the system, it does not propagate but remains there forever frozen. Such excitations have been studied in the literature \cite{thorpe1995bulk, sun2012surface, kane2014topological} and are known as floppy modes.
It is interesting to note that these solutions have a hidden holomorphicity property related to the fact that, when $\sigma = -1$, the boundary conditions \eqref{BC} become the Cauchy-Riemann equations for the field $u_x + i u_y$, see Appendix \ref{AppC} for more details. 

\section{Conclusions and outlook}
We analyzed  Rayleigh edge waves that travel on the edge of two-dimensional crystals in the presence of Lorentz forces and mapped out how their propagation direction depends on the Poisson ratio, see Fig. \ref{Fig:regimes}. The existence of these waves is not protected by topology, but rather originates from spontaneously broken translational symmetry.
In addition to skyrmion crystals, we expect our findings to be directly applicable to boundary excitations of screened Wigner crystals in an external magnetic field \cite{edge_vignale}. Moreover, our results shed new light on elastic gyroscopic systems \cite{brun2012vortex}, where edge modes are currently under active investigation \cite{nieves2020rayleigh, zhao2020non}. Our work indicates that in all these systems the chirality of Rayleigh edge waves can be controlled by changing the elastic properties of the medium.

Extensions of this study to Abrikosov vortex crystals in superconductors and superfluids \cite{Sonin2016} are non-trivial exciting frontiers.
It would be also intriguing to generalize this work and investigate edge excitations in two-dimensional crystals, where time-reversal breaking originates from a different mechanism, such as for example the odd elasticity discovered in \cite{scheibner2020odd, banerjee2020active}.

\begin{acknowledgements}
We would like to acknowledge the useful discussions held with Christian Back, Umberto Borla, Luca Delacr\'etaz, Markus Garst,  Steven Girvin, Nicola Pancotti, Christian Pfleiderer, Frank Pollmann, Leo Radzihovsky, Cosimo Rusconi, Anton Souslov and Oleg Tchernyshyov.
Our work is funded by the Deutsche Forschungsgemeinschaft (DFG, German Research Foundation) under Emmy Noether Programme grant no.~MO 3013/1-1 and under Germany's Excellence Strategy - EXC-2111 - 390814868.
\end{acknowledgements}

\appendix


\begin{widetext}

\section{Analytic values of $\sigma_1$ and $\sigma_2$, edge-wave dispersions at $\sigma=1/3$, $\sigma=\sigma_{1,2}$ and floppy modes} \label{AppA}
Insertion of the edge-wave ansatz into the stress-free boundary condition \eqref{BCS} results in an equation for $\alpha_\pm$:
\begin{eqnarray}
\scalebox{0.8}{%
$\frac{2 (\sigma -1) \left(\alpha _+^2 \left(\sigma  \left(-\sqrt{\frac{1}{1-\sigma }} \sqrt{2-2 \alpha _+^2}+2 \sqrt{1-\alpha _+}-2 \sqrt{\alpha _++1}\right)+\sqrt{\frac{1}{1-\sigma }} \sqrt{2-2 \alpha _+^2}-2 \sqrt{2} \sqrt{\frac{1}{1-\sigma }} \sigma ^2\right)+(\sigma +1)^2 \left(\sqrt{1-\alpha _+}-\sqrt{\alpha _++1}\right)+(\sigma +1)^2 \alpha _+ \left(\sqrt{1-\alpha _+}+\sqrt{\alpha _++1}\right)\right)}{(\sigma -1)^2 \alpha _+^2-(\sigma +1)^2}=0$} \label{eq:alphaplusminus_1}\\
\scalebox{0.8}{%
$\frac{2 (\sigma -1) \left(\alpha _-^2 \left(\sigma  \left(\sqrt{\frac{1}{1-\sigma }} \sqrt{2-2 \alpha _-^2}+2 \sqrt{1-\alpha _-}-2 \sqrt{\alpha _-+1}\right)-\sqrt{\frac{1}{1-\sigma }} \sqrt{2-2 \alpha _-^2}+2 \sqrt{2} \sqrt{\frac{1}{1-\sigma }} \sigma ^2\right)+(\sigma +1)^2 \left(\sqrt{1-\alpha _-}-\sqrt{\alpha _-+1}\right)+(\sigma +1)^2 \alpha _- \left(\sqrt{1-\alpha _-}+\sqrt{\alpha _-+1}\right)\right)}{(\sigma -1)^2 \alpha _-^2-(\sigma +1)^2}=0$} \label{eq:alphaplusminus_2}
\end{eqnarray}
We are considering non-negative values of $\omega$, thus $\alpha_\pm\geq 0$.  The form of the spectrum \eqref{Disp} yields for $\kappa$ given by \eqref{kappas} the values
\begin{eqnarray}
\kappa_{1,2} = \sqrt{1\pm \alpha} |k|.
\end{eqnarray}
In order to have both $\kappa_{1,2}$ real, the condition $\alpha \leq 1$ must be satisfied. The analytic values of $\sigma_1$ and $\sigma_2$ can be found by imposing these limits. We first note that for $\alpha_+ \rightarrow 1$ one finds $\sigma \rightarrow 0$ using Eq. \eqref{eq:alphaplusminus_1}, and  thus 
\begin{equation}
\sigma_1 = 0.
\end{equation}

The value $\sigma_2$ is obtained by letting $\alpha_- \rightarrow 1$ in Eq. \eqref{eq:alphaplusminus_2}. In this limit that equation reduces to
\begin{eqnarray}
\sigma_2 \sqrt{1-\sigma_2 }  -1 + \sigma_2  = 0
\end{eqnarray}
with solution
\begin{eqnarray}
\sigma_2 = \frac{\sqrt{5}-1}{2}\equiv \varphi^{-1},
\end{eqnarray}
which is the inverse of the golden ratio $\varphi$.

\subsection{Symmetric point $\sigma = 1/3$}
When $\sigma = 1/3$, both equations \eqref{eq:alphaplusminus_1} and \eqref{eq:alphaplusminus_2} reduce to the same form, thus $\alpha_+ = \alpha_-$. The equation that is satisfied by $\alpha_\pm$ is
\begin{eqnarray}
\left(-3 \sqrt{3-3 \alpha ^2}+3 \sqrt{1-\alpha }-3 \sqrt{\alpha +1}+\sqrt{3}\right) \alpha ^2+8 \left(\sqrt{1-\alpha }+\sqrt{\alpha +1}\right) \alpha +8 \left(\sqrt{1-\alpha }-\sqrt{\alpha +1}\right) = 0.
\end{eqnarray}
It is straightforward to verify that the only admissible solution is 
\begin{eqnarray}
\alpha_\pm = \frac{2 \sqrt{2}}{3}.
\end{eqnarray}
Thus the long-wavelength edge-wave dispersion at $\sigma = 1/3$ takes on the particularly simple form
\begin{eqnarray}
\omega = 2\sqrt{\frac{2}{3}}   \frac{v_2}{\Omega} k^2.
\end{eqnarray}
\subsection{Asymptotic behaviour of $\alpha_\pm$ at $\sigma = \pm 1$}
When $\sigma  \rightarrow 1^-$ the value of the corresponding $\alpha_+$ tends to 0. By setting 
\begin{eqnarray}
\sigma &=& 1-\epsilon\\
\alpha &=& \delta
\end{eqnarray}
in equation \eqref{eq:alphaplusminus_1} and expanding in small $\delta, \epsilon$, we arrive at the equation
\begin{eqnarray}
2 \delta  \epsilon -\sqrt{2} \delta ^2 \sqrt{\epsilon } = 0
\end{eqnarray}
which has the solution $\delta = \sqrt{2} \sqrt{\epsilon}$. This yields the asymptotic
\begin{eqnarray}
\alpha_+ \sim \sqrt{2} \sqrt{1-\sigma}  \text{ as } \sigma \rightarrow 1^{-}
\end{eqnarray}
and as a consequence
\begin{eqnarray}
\omega \sim \frac{2 v_2}{\Omega} k^2  \text{ as } \sigma \rightarrow 1^{-}.
\end{eqnarray}
When $\sigma \rightarrow -1^+$, the value of $\alpha_-$ tends to $0$. Here we set
\begin{eqnarray}
\sigma &=& -1+\epsilon\\
\alpha &=& \delta
\end{eqnarray}
and upon expanding \eqref{eq:alphaplusminus_2} we find $\delta = {3\epsilon}/{2} $ and thus
\begin{eqnarray}
\alpha_-\sim \frac{3}{2} (1+\sigma) \text{ as } \sigma\rightarrow -1^+
\end{eqnarray}
and therefore
\begin{eqnarray}
\omega \sim \frac{3 v_2}{2\Omega} (1+\sigma) k^2  \text{ as } \sigma \rightarrow -1^{+}.
\end{eqnarray}
Since $\sigma \rightarrow -1^+$ is equivalent to $v_1 \rightarrow 0$, we can also write this asymptotic relation as
\begin{eqnarray}
\omega \sim \frac{6v_1}{\Omega}  k^2  \text{ as } v_1 \rightarrow  0.
\end{eqnarray}
In both limits,  $\sigma \rightarrow -1^{+}$ and $\sigma \rightarrow 1^{-}$, the spectra become flat. Such flat spectra are associated with excitations called floppy modes. 

\subsection{Floppy modes at $\sigma=-1$}
As discussed in the main text, for $\sigma = -1$ the system supports floppy modes. Setting $\sigma=-1$ and inserting the edge-wave ansatz \eqref{superposition}  into the equations of motion \eqref{EOM} produces two modes with $\kappa_\mp= \sqrt {k^2 - \left(\omega^2  \pm \omega \Omega\right)/v_2}$ and circular polarizations $\epsilon_\mp = \pm i$. The boundary conditions \eqref{BC} enforce $\omega = 0$ or $\omega = \Omega$ and $\kappa_\mp= |k|$. The latter is automatically satisfied for the $\omega = 0$ bulk mode. But, for $\omega = \Omega$, the $\kappa_+$-mode violates this condition. 

For $\omega = 0$ we find the floppy mode
\begin{equation}\label{eg:gapplessC0}
\textbf{u} =\left( -i~\text{sign}(k),1
\right)^{\rm {T}}e^{ikx+|k|y},
\end{equation}
while the time-dependent solution with $\omega = \Omega$ only exists for $k>0$ and has the form 
\begin{equation}\label{eg:gappedC0}
\textbf{u} =\left( -i ,1
\right)^{\rm {T}}e^{i(k x - \Omega t)+k y}.
\end{equation}
We assumed above that $\Omega > 0$. If, instead, $\Omega < 0$, then the time-dependent solution has the frequency $\omega = -\Omega$. This change of sign modifies the sign of the allowed $k$ values in Eq. \eqref{eg:gappedC0} and thereby reverses the direction of propagation.
\section{Finite-Difference Solution of the Equations of Motion} \label{AppB}
In the main part of the paper, we displayed snapshots of Rayleigh waves propagating along the boundary of a square-grid system. These snapshots are obtained from finite-difference simulations of the partial differential equation (\ref{EOM}) subject to the boundary conditions (\ref{BC}). To this end, we discretize space by introducing a quadratic grid. Along the vertical sides of the square, we use periodic boundary conditions. In the horizontal direction, where we observe surface waves, we use the free boundary conditions (\ref{BC}). In the absence of the magnetic field term, our bulk equations of motion are reduced to those considered in the classic finite-difference treatment of Kelly et al.\cite{kelly1976synthetic}, where an explicit scheme was introduced. We employ the same discretization, but treat the magnetic field term exactly. \newline
The equations of motion are discretized after rewriting them as first-order equations in time, by introducing the velocity fields $w_x = \dot{u}_x$ and $w_y = \dot{u}_y$:
\begin{align*}
[w_{x}]^{n,m}_{l+1} & = \cos(h \Omega)[w_{x}]^{n,m}_{l} + \sin(h \Omega)[w_{y}]^{n,m}_{l}+h[F_{x}]^{n,m}_l\\
[w_{y}]^{n,m}_{l+1} & = \cos(h \Omega)[w_{y}]^{n,m}_l - \sin(h \Omega)[w_{x}]^{n,m}_{l}+h[F_{y}]^{n,m}_l \\
[u_{x}]^{n,m}_{l+1} & =[u_{x}]^{n,m}_l+h [w_{x}]^{n,m}_l \\
[u_{y}]^{n,m}_{l+1} & =[u_{y}]^{n,m}_l+h [w_{y}]^{n,m}_l
\end{align*}
where the subscript $l$ is the time index, $h$ is the discrete time step  and $F_x$ and $F_y$ are the centrally discretized elastic forces.  The magnetic field discretization is exact in the absence of elastic forces. \newline
Since our focus is on Rayleigh waves, the boundary is particularly important. We use a stable finite-difference scheme that was invented by Vidale and Clayton \cite{vidale1986stable} for the study of surface waves. In their method, an auxiliary horizontal layer is added to the last grid layer, and the actual free surface is considered to be in-between these two layers. The time evolution in all but the additional layer is carried out by the discretized bulk equations of motion. The updates on the added layer are derived from the boundary conditions, which are
\begin{align*}
\partial_{y}u_{x}+\partial_{x}u_{y} & =0\\
\sigma \partial_{x}u_{x}+\partial_{y}u_{y} & =0.
\end{align*}
These conditions have to be imposed at the free surface, which is obtained from the last two layers by averaging. Discretizing these equations using central differences yields
\begin{align*}
u_{x}^{1,n}-u_{x}^{0,n}+\frac{1}{2}\left[\frac{u_{y}^{0,n+1}+u_{y}^{1,n+1}}{2}-\frac{u_{y}^{0,n-1}+u_{y}^{1,n-1}}{2}\right] & =0\\
\frac{\sigma}{2}\left[\frac{u_{x}^{0,n+1}+u_{x}^{1,n+1}}{2}-\frac{u_{x}^{0,n-1}+u_{x}^{1,n-1}}{2}\right]+(u_{y}^{1,n}-u_{y}^{0,n}) & =0,
\end{align*}
where the indices $0$ and $1$ denote the last and penultimate horizontal layers, respectively. These equations have to be solved in order to find the $u_x^{0,n}$ and $u_y^{0,n}$.
We can cast these equations as matrix equations by introducing the tridiagonal matrix $T$ with components $T_{nm}=\delta_{n,m-1}-\delta_{n,m+1}$ and forming vectors $\mathbf u_x^{0}$ and $\mathbf u_x^{1}$ out of the displacements: 
\begin{eqnarray} \label{eq:coupledMatr}
\frac{1}{4}T\mathbf{u}_{y}^{0}-\mathbf{u}_{x}^{0} & =&-\frac{1}{4}T\mathbf{u}_{y}^{1}-\mathbf{u}_{x}^{1}\\
\frac{\sigma}{4}T\mathbf{u}_{x}^{0}-\mathbf{u}_{y}^{0} & =&-\frac{\sigma}{4}T\mathbf{u}_{x}^{1}-\mathbf{u}_{y}^{1}
\end{eqnarray}
The right-hand sides are given. Solving the first equation for $\mathbf{u}_{x}^{0}$ and inserting it into
the second, we find an equation for $\mathbf{u}_{y}^{0}$ alone:
\begin{align*}
\left(\mathbb{I}-\frac{\sigma}{16}T^{2}\right)\mathbf{u}_{y}^{0} & =\frac{\sigma}{16}T^{2}\mathbf{u}_{y}^{1}+\frac{\sigma}{2}T\mathbf{u}_{x}^{1}+\mathbf{u}_{y}^{1}
\end{align*}
The first step is to solve this matrix equation for $\mathbf{u}_{y}^{0}$.  In the second step, one finds $\mathbf{u}_{x}^{0}$ by using  equation (\ref{eq:coupledMatr}):
\begin{eqnarray}
\mathbf{u}_{x}^{0}=\frac{1}{4}T\mathbf{u}_{y}^{0}+\frac{1}{4}T\mathbf{u}_{y}^{1}+\mathbf{u}_{x}^{1} \label{eq:particularMatrixProblem}
\end{eqnarray}
The matrix equation in the first step is of the form
\begin{eqnarray}
\left(\mathbb{I}+aT^{2}\right)\mathbf{x}=\mathbf{b}. \label{eq:genMatrixProblem}
\end{eqnarray}
As noted in \cite{vidale1986stable} this matrix is pentadiagonal and can be solved by methods similar to those for tridiagonal matrices \cite{claerbout1985fundamentals}. Written out, this matrix equation becomes
\[
ax_{n-2}+(1-2a)x_{n}+ax_{n+2}=b_{n},
\]
which is a three-term recursion relation that only connects the even/odd indexed terms. It can be solved by making a two-term recursion ansatz
\begin{equation}
x_{n}=A_{n}x_{n+2}+B_{n}\label{eq:two-term}.
\end{equation}
We use this to eliminate $x_{n+2}$ in the three-term recursion, which  results in
\begin{align*}
x_{n} & =-\frac{a}{1-2a+aA_{n-2}}x_{n-2}+\frac{b_{n}-aB_{n-2}}{1-2a+aA_{n-2}}.
\end{align*}
Comparing this with the two-term ansatz, we find:
\begin{align}
A_{n} & =-\frac{a}{1-2a+aA_{n-2}}\label{eq:aeq}\\
B_{n} & =\frac{b_{n}-aB_{n-2}}{1-2a+aA_{n-2}}\label{eq:beq}
\end{align}
Let us assume that the initial conditions $x_{0},x_{1}$ are specified and put $A_{0}=0,B_{0}=x_{0}$ and $A_{1}=0,B_{1}=x_{1}$. Then we can determine from (\ref{eq:aeq})-(\ref{eq:beq}) all the remaining $A_{n},B_{n}$. Next we take the initial condition on $x_{N}$ and use (\ref{eq:two-term}) to determine all the $x_{i}$ between $i=1$ and $i=N-1$, thereby solving the inversion problem (\ref{eq:genMatrixProblem}). In our particular problem (\ref{eq:particularMatrixProblem}) we have 
\begin{align*}
a & =-\frac{\sigma}{16}\\
b_{n} & =\left[\frac{\sigma}{16}T^{2}\mathbf{u}_{y}^{1}+\frac{\sigma}{2}T\mathbf{u}_{x}^{1}+\mathbf{u}_{y}^{1}\right]_{n}.
\end{align*}
A full update step consists of a bulk update followed by the boundary updates of auxiliary layers (on the top and bottom of the square).

\section{Complex Formulation of the Equations of Elasticity and Holomorphicity at $\sigma = -1$}\label{AppC} 
When the Poisson ratio takes on the value $\sigma=-1$, the edge-wave solutions have a hidden property. To see this we reformulate the elasticity equations in complex form by combining the real strain components $u_x$ and $u_y$ into one complex field $\psi \equiv u_x + i u_y$. The equations of motion \eqref{EOM} are 
\begin{align*}
\ddot{u}_{x}+\Omega\dot{u}_{y}-2v_{1}\partial_{x}(\partial_{x}u_{x}+\partial_{y}u_{y})-v_{2}(\partial_{x}^{2}+\partial_{y}^{2})u_{x} & =0\\
\ddot{u}_{y}-\Omega\dot{u}_{x}-2v_{1}\partial_{y}(\partial_{x}u_{x}+\partial_{y}u_{y})-v_{2}(\partial_{x}^{2}+\partial_{y}^{2})u_{y} & =0
\end{align*}
and by multiplying the second equation by $i$ and adding it to the first, we obtain
\begin{equation}
\ddot{\psi}-i\Omega\dot{\psi}-4v_{1}(\partial_{\bar{z}}^{2}\bar{\psi}+\partial_{z}\partial_{\bar{z}}\psi)-4v_{2}\partial_{z}\partial_{\bar{z}}\psi  =0,
\end{equation}
where we introduced the complex derivatives $\partial_{z} \equiv \left({\partial_{x}-i\partial_{y}}\right)/{2}$ and $\partial_{\bar{z}}  \equiv \left({\partial_{x}+i\partial_{y}}\right)/{2}$. The boundary conditions (\ref{BC}) in real space read
\begin{eqnarray}
\partial_{x}u_{y}+\partial_{y}u_{x} & =0 \label{eq:BC_real_1} \\
\sigma\partial_{x}u_{x}+\partial_{y}u_{y} & =0. \label{eq:BC_real_2}
\end{eqnarray}
Multiplying the second equation by $i$ and adding it to the first, we obtain the boundary conditions in complex form
\begin{equation}
\label{eq:BC_complex}
(3-\sigma)\partial_{\bar{z}}\psi =(1+\sigma)(\partial_{\bar{z}}\bar{\psi}+\partial_{z}\bar{\psi} + \partial_{z}\psi).
\end{equation}
 At $\sigma = -1$ the boundary conditions \eqref{eq:BC_real_1} and \eqref{eq:BC_real_2} are the Cauchy-Riemann equations for the real and imaginary parts of $\psi$ at $y = 0$.  In addition, the real parts of the modes (\ref{eg:gapplessC0}) and (\ref{eg:gappedC0}) give rise to $\psi$'s that are holomorphic functions of the complex variable $z \equiv x + i y$ in the bulk. In particular, the time-independent mode yields $\psi   =  i \exp (-i |k| z)$, while the time-dependent mode is $\psi = i \exp(-ik z + i\Omega t)$. Using a conformal transformation we can map these edge-modes, which are localized near the boundary of the complex half-plane, onto edge-waves that propagate along the boundary of an arbitrarily shaped region. In other words, the transformed solutions will satisfy the boundary conditions on the new edge and solve the bulk (Laplace) equations of motion \cite{needham1998visual}. 
\section{Symmetric Edge Spectrum} \label{AppD}

In Appendix \ref{AppA}, we have shown that a symmetric spectrum of edge excitations emerges for the value of Poisson ratio $\sigma=1/3$, i.e. for equal elastic moduli, $C_1=C_2\equiv C$.
Hereby we show that this property holds even at NLO, see Figure \ref{Fig:equal}.

\begin{figure}
\includegraphics[width=0.5\columnwidth]{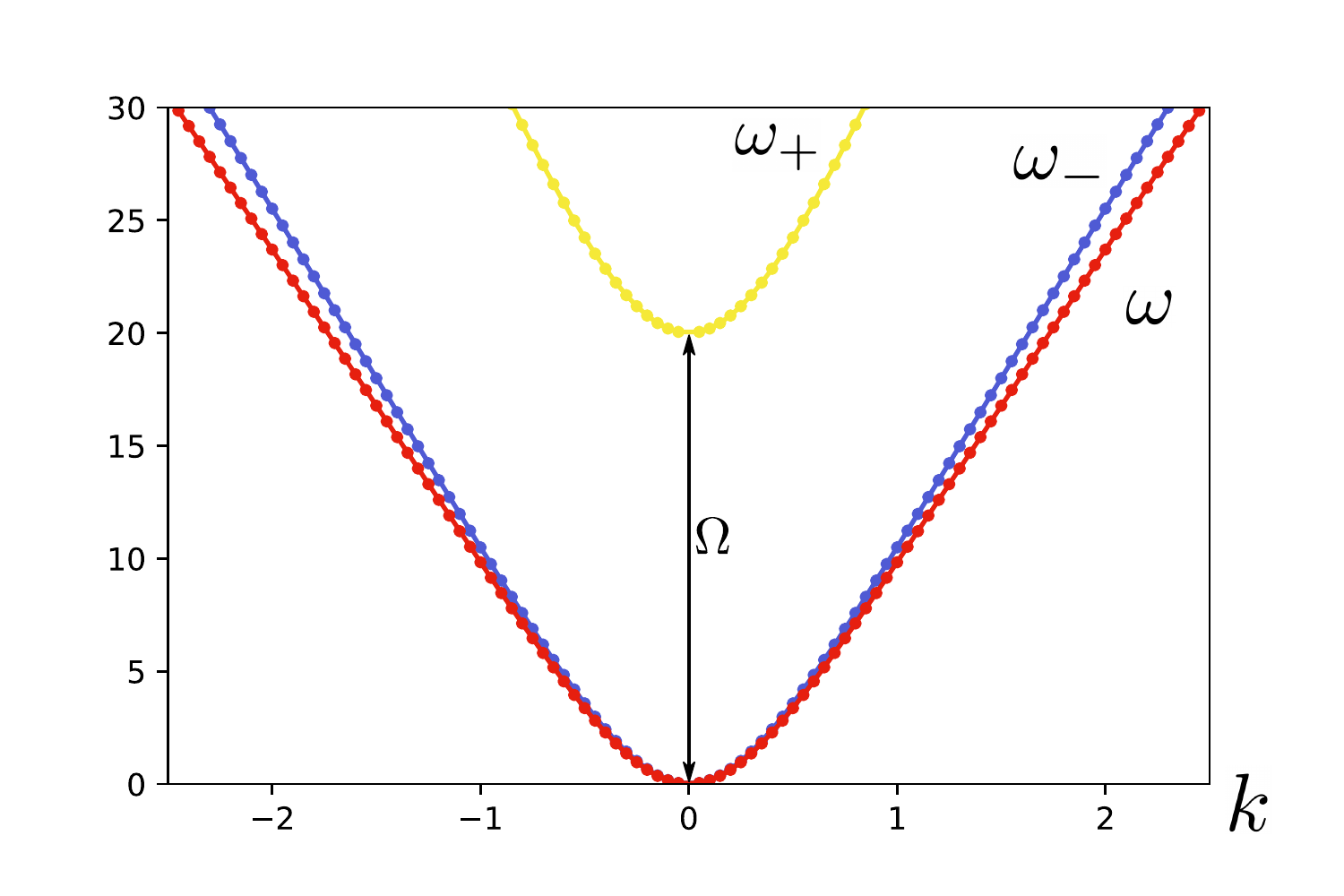} 
\caption{Numerical solution of the edge wave dispersion (red) for equal elastic moduli $C_1 = C_2$. For comparison, also the dispersions relations of the bulk modes $\omega_-$ (blue) and $\omega_+$ (yellow) are plotted.}
\label{Fig:equal}
\end{figure} 
The proof is based on the characteristic equation $d(k, \omega)=0$ of the matrix appearing in the boundary conditions (\ref{BCS}).

The expressions are cumbersome and it turns out to be more convenient to study the parity property of the auxiliary function $\tilde{d}\left(k,\omega\right) \equiv d\left(k,\omega\right)/\left[\epsilon_+\left(k,\omega\right) - \epsilon_-\left(k,\omega\right)\right]$ instead of the characteristic polynomial $d\left(k,\omega\right)$.  In terms of the inverse decay lengths $\kappa_{\mp}\left( k,\omega \right)$ and the polarizations $\epsilon_{\mp}\left( k,\omega \right)$ the auxiliary function takes the following form
\begin{equation}\label{dtilde}
\tilde{d} \left(k,\omega\right)= \left(k^2+3 \kappa_- \kappa_+ \right)+ \left(\kappa_+ - \kappa_- \right) \left[\underbrace{\frac{\epsilon_- \epsilon_+-3}{\epsilon_+-\epsilon_-}\left(ik\right)}_{\equiv g\left(k,\omega\right)} \right].  
\end{equation}
We will argue that the auxiliary function $\tilde{d}$ is an even function of the wavevector $k$.
First, after introducing $2C/\rho \equiv v$, we notice that
\begin{equation}
\kappa_{\mp} \left(k,\omega \right) = \sqrt{k^2- \left(2 \omega ^2\pm  \omega   \sqrt{3 \Omega^2+\omega ^2}\right)/3
   v}
\end{equation}
are even functions of $k$. As a result, all the functions outside the square brackets in (\ref{dtilde}) are even. 
Polarization functions have no parity symmetry
\begin{equation*}
\epsilon_{\mp}\left(k,\omega\right)=  - i \frac{3\Omega^2\omega + 2k \sqrt{3v \left(-2\omega^2 \mp \sqrt{3\Omega^2 \omega^2+\omega^4}+3v k^4\right)}}{\omega^2-\sqrt{3\Omega^2 \omega^2+\omega^4}-6v k^2},
\end{equation*} 
however, together with $(ik)$, they lead to a function inside the square brackets
\begin{equation}
g\left(k,\omega\right) = \frac{-k^2}{{\sqrt{\omega^2 \Omega^2 + \omega^4/3}}}\left[  \sqrt{-v \left(2\omega^2 + \sqrt{3\omega^2 \Omega^2 + \omega^4}-3v k^2\right)}  +  \sqrt{-v \left(2\omega^2 - \sqrt{3 \omega^2 \Omega^2  + \omega^4}- 3v k^2\right)} \right]
\end{equation}
which is manifestly even under the change of sign of the wavevector. This proves that $\tilde{d}\left(k,\omega\right)= \tilde{d}\left(-k,\omega\right)$ therefore the edge-wave spectrum at $\sigma=1/3$ is symmetric, $\omega(k)= \omega(-k)$.

\end{widetext}

\bibliography{biblio}

\end{document}